\documentclass[showpacs,preprintnumbers,amsmath,amssymb]{revtex4}

\usepackage{graphicx}
\usepackage{dcolumn}
\usepackage{bm}
\begin{document}
\title{Weak Pion Production From Nuclei}
\author{S. K. Singh, M. Sajjad Athar and Shakeb Ahmad}
\affiliation{Department of Physics, Aligarh Muslim University, Aligarh-202 002, India.}
\date{\today}
 \begin{abstract}
 The charged current pion production induced by neutrinos in $^{12}{C}$, $^{16}{O}$ and $^{56}{Fe}$ nuclei has been studied. The calculations have been done for the coherent as well as the incoherent processes assuming $\Delta$ dominance and take into account the effect of Pauli blocking, Fermi motion and the renormalization of $\Delta$ in the nuclear medium. The pion absorption effects have also been taken into account.
\end{abstract}
\pacs{25.30Pt,23.40.Bw,13.15.+g}
\maketitle 
\section{Introduction}
The pion production processes from nucleons and nuclei at intermediate energies are important tools to study the hadronic structure. The dynamic models of the hadronic structure are used to calculate the various nucleon and transition form factors which are tested by using the experimental data on photo, electro and weak pion production processes on nucleons. The weak pion production along with the electro pion production from nucleon in the $\Delta$-resonance region is used to determine the various electroweak $N-\Delta$ transition form factors. The neutrino induced pion production experiments performed at CERN\cite{cern}, ANL\cite{barish} and BNL\cite{kitagaki} laboratories have been analyzed to obtain informations on these form factors. The early neutrino experiments performed at CERN[1] have low statistics and are done using heavy nuclear targets and their analysis have uncertainties related to the nuclear corrections. The later experiments performed at ANL\cite{barish} and BNL\cite{kitagaki} are done in hydrogen and deuterium and are free from nuclear medium corrections. These experiments also have high statistics and provide reasonable estimates of the dominant form factors in $N-\Delta$ transitions. However, with the availability of the new neutrino beams in intermediate energies at K2K\cite{k2k} and MiniBooNE\cite{boone}, it is desirable that various $N^\star-N$ weak transition form factors are determined for low lying nuclear resonances like $\Delta(1232)$, $N^*(1440)$, $N^*(1535)$, etc. There is a considerable activity in this field, specially, in the determination of electromagnetic transition form factors using the photo and electroproduction data from MAINZ, BONN and TJNAF laboratories\cite{mainz}. It is desirable that such attempts be extended to the determination of weak transition form factors also. 
\begin{figure}[h]
\includegraphics{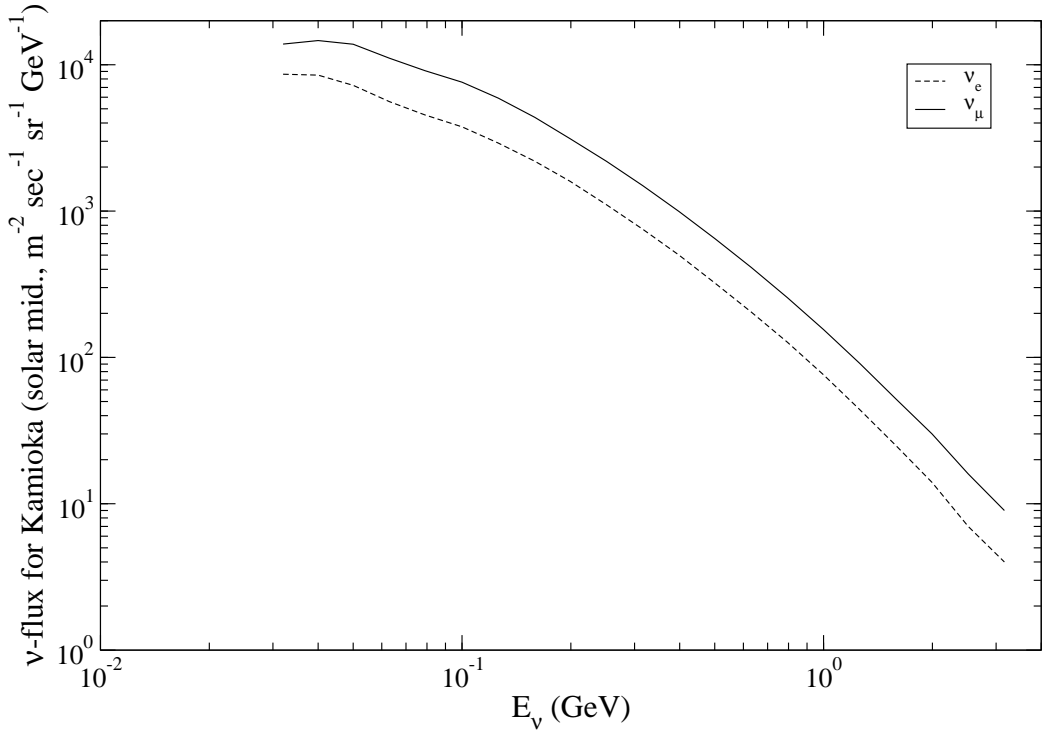}
\caption{The atmospheric $\nu$-flux for Kamioka site[7]}
\end{figure}
Recently, the weak pion production processes have become very important in the analysis of the neutrino oscillation experiments with atmospheric neutrinos. The energy spectrum of atmospheric neutrino at Kamioka site\cite{honda}, shown in Fig.1 is such that the weak pion production contributes about $20\%$ of the quasielastic lepton production and is a major source of uncertatinty in the identification of electron and muon events. In particular, the neutral current $\pi^0$ production contributes to the background of $e^\pm$ production while $\pi^\pm$ production contributes to the background of $\mu^\pm$ production. This is because both the particles i.e. $\pi^0$ and $e^\pm$ are identified through the detection of photons and $\pi^\pm$ and $\mu^\pm$ are identified through single track events in the detection of neutrino oscillation experiments. Moreover, the neutral current $\pi^0$ production plays very important role in distinguishing between the two oscillation mechanisms $\nu_\mu \rightarrow \nu_\tau$ and $\nu_\mu \rightarrow \nu_s$\cite{vissani}.

The neutrino oscillation experiments are generally performed with detectors which use material with nuclei like $^{12}{C}$, $^{16}{O}$, $^{56}{Fe}$, etc. as targets. It is therefore desirable that nuclear medium effects be studied in the production of leptons and pions induced by the atmospheric as well accelerator neutrinos used in these neutrino oscillation experiments. 

There have been many attempts in past to calculate the weak pion production\cite{adler}, but very few attempts have been made to estimate the nuclear effects and their influence on the weak pion production in nuclei\cite{adler1}. The recent experiments on neutrino oscillation experiments with atmospheric and accelerator neutrinos in the intermediate energy region have started a fresh interest in the study of weak pion production of pions from nuclei\cite{kim}-\cite{ruso}. We report in this paper the results of a calculation of the weak pion production in nucleus assuming $\Delta$ dominance. The effect of Pauli blocking, Fermi motion and renormalisation of weak $\Delta$ properties in the nuclear medium are taken into account in a local density approximation following the methods of Ref.\cite{ruso}. 

In section-2, we describe the matrix elements for the production of $\Delta$ resonance from free nucleons using the information about the $N-\Delta$ transition form factors as determined from experiments. These matrix elements are then used to calculate the weak pion production in nuclei. We have described and discussed the incoherent weak production of pions in section-3 and the coherent weak pion production in section-4 with conclusion given in section-5. 

\section{Weak Pion Production from Nucleons}
The weak pion production processes like 
\begin{equation}
\nu_l(\bar\nu_l) + N\rightarrow l^{-}(l^+) + N + \pi^\alpha
\end{equation}
and
\begin{equation}
\nu_l(\bar\nu_l) + N\rightarrow \nu_l(\bar\nu_l) + N + \pi^\beta
\end{equation}
where N can be a proton or neutron and $\pi^\alpha(^\beta)$ are the different possible charged states ($\pi^+,\pi^-$ and $\pi^0$) of the pions produced and are determined by the lepton number and charge conservation in the charged current and neutral current reactions in Eqns.(1) and (2) respectively.

The pion production induced by charged current neutrino interaction is calculated in the standard model of electroweak interactions using the Lagrangian given by 
\begin{equation}
{\it L} = \frac{G_F}{\sqrt 2} l_\mu(x) ~ {J^{\mu}(x)}^W, where
\end{equation}
$l_\mu(x)=\bar{\psi}(k^\prime)\gamma_\mu(1-\gamma_5)\psi(k)$ and 
${J^{\mu}(x)}^W$ is the hadronic current given by ${J^{\mu}(x)}^W=cos{\theta_c}~(V^\mu(x)+ A^\mu(x) + ~h.c.)$ for the charged current reactions .

The matrix element of the hadronic current ${J^\mu}^W$ is generally calculated using the nucleon and meson exchanges and the resonance excitation diagrams. However, it has been shown that in the intermediate energy region of about 1GeV, the dominant contribution to the pion production from nucleons as well as from nuclei comes from the $\Delta$ resonance due to very strong P wave pion nucleon coupling leading to $\Delta$ resonance. Furthermore, the angular distribution and the energy distribution of the pions is dominated by the $\Delta$ contribution while the other diagrams contribute to the tail region due to the interference of the nucleon and meson exchange diagrams with the $\Delta$ resonance diagram\cite{paschos}. 

In this model the matrix element for the neutrino excitation of the $\Delta$ resonance through charged current reaction i.e. 
\begin{equation}
\nu_l(k)+p(p)\rightarrow l^{-}(k^\prime)+\Delta^{++}(p^\prime) and
\end{equation}
\begin{equation}
\nu_l(k)+n(p)\rightarrow l^{-}(k^\prime)+\Delta^{+}(p^\prime)
\end{equation}
is written as 
\begin{equation}
T = \frac{G_F}{\sqrt{2}}\cos{\theta_c} l_\mu ~{(V^{\mu} + A^{\mu})}
\end{equation}
where $l_\mu$ is the leptonic current and $V^\mu$ and $A^\mu$ defines the vector and axial vector part of the transition hadronic current between $N$ and $\Delta$ states for the charged current interaction.  The most general form of the matrix elements of hadronic currents between the p and $\Delta^{++}$ states, in Eqn.4 are given by\cite{barish},\cite{adler},\cite{ruso}:
\begin{eqnarray}
<\Delta^{++}|V^\mu|p>&=&{\sqrt 3}\bar{\psi}_\alpha(p^\prime)\left[\left(\frac{C^V_{3}(q^2)}{M}(g^{\alpha\mu}{\not q}-q^\alpha{\gamma^\mu})\right.\right.\nonumber\\
&+&\frac{C^V_{4}(q^2)}{M^2}(g^{\alpha\mu}q\cdot{p^\prime}-q^\alpha{p^{\prime\mu}})\nonumber\\
&+&\left.\left.\frac{C^V_5(q^2)}{M^2}(g^{\alpha\mu}q\cdot p-q^\alpha{p^\mu})+\frac{C^V_6(q^2)}{M^2}q^\alpha q^\mu\right)\gamma_5\right]u(p)
\end{eqnarray}
\begin{eqnarray}
<\Delta^{++}|A^\mu|p>&=&{\sqrt 3}\bar{\psi}_\alpha(p^\prime)\left[\left(\frac{C^A_{3}(q^2)}{M}(g^{\alpha\mu}{\not q}-q^\alpha{\gamma^\mu})+\frac{C^A_{4}(q^2)}{M^2}(g^{\alpha\mu}q\cdot{p^\prime}-q^\alpha{p^{\prime\mu}})\right.\right.\nonumber\\ 
&+&\left.\left.C^A_{5}(q^2)g^{\alpha\mu}+\frac{C^A_6(q^2)}{M^2}q^\alpha q^\mu\right)\right]u(p)
\end{eqnarray}
Here ${\psi_\alpha}(p^\prime)$ and u(p) are the Rarita Schwinger and Dirac spinors for $\Delta$ and nucleon of momenta $p^\prime$ and $p$ respectively, $q(=p^\prime-p=k-k^\prime)$ is the momentum transfer and $C^V_i$(i=3-6) are vector and $C^A_i$(i=3-6) are axial vector transition form factors. The vector form factors $C^V_i$(i=3-6) are determined by using the conserved vector current(CVC) hypothesis which gives $C_6^V(q^2)$=0 and relates $C_i^V$(i=3,4,5) to the electromagnetic form factors which are determined from photoproduction and electroproduction of $\Delta$'s. Using the analysis of these experiments\cite{paschos},\cite{dufner} we take for the vector form factors
\begin{eqnarray}
C_{5}^V=0,\;\;\;\;C_{4}^V=-\frac{M}{M_\Delta}C_{3}^V,~~~~~~~~~\mbox{and}\nonumber\\
C_{3}^V(q^2)=\frac{2.05}{(1-\frac{q^2}{M_V^2})^2}, ~M_V^2=0.54GeV^2
\end{eqnarray}
The axial vector form factor $C_{6}^A(q^2)$ is related to ${C_5}^A(q^2)$ using PCAC and is given by
\begin{equation}
C_{6}^A(q^2)={C_5}^A(q^2)\frac{M^2}{{m_\pi}^2-q^2}
\end{equation}
The remaining axial vector form factor ${C^{A}_{i=3,4,5}}(q^2)$ are taken from the experimental analysis of the neutrino experiments producing $\Delta$'s in proton and deuteron targets\cite{barish}-\cite{kitagaki}. These form factors are not uniquely determined but the following parameterizations give a satisfactory fit to the data.
\begin{equation}
{C^{A}_{i=3,4,5}}(q^2)=C_{i}^A(0)\left[1+\frac{{a_i}q^2}{b_i-q^2}\right]\left(1-\frac{q^2}{{M_A}^2}\right)^{-2}
\end{equation}
with $C_{3}^A(0)=0, C_{4}^A(0)=-0.3,C_{5}^A(0)=1.2, a_4=a_5=-1.21,\;\; b_4=b_5=2GeV^2$, $M_A=1.28GeV$.
Using the hadronic current given in Eqns.7 and 8, the energy spectrum of the outgoing leptons is given by
\begin{equation}
\frac{d^2\sigma}{dE_{k^\prime}d\Omega_{k^\prime}}=\frac{1}{8\pi^3}\frac{1}{MM^\prime}\frac{k^\prime}{E_\nu}\frac{\frac{\Gamma(W)}{2}}{(W-M^\prime)^2+\frac{\Gamma^2(W)}{4.}}L_{\mu\nu}J^{\mu\nu}  
\end{equation}
where $W=\sqrt{(p+q)^2}$ and $M^\prime$ is mass of $\Delta$,
\begin{eqnarray}
L_{\mu\nu}&=&{\bar\Sigma}\Sigma{l_\mu}^\dagger l_\nu=L_{\mu\nu}^S + iL_{\mu\nu}^A\nonumber\\
&=&k_\mu k_\nu^\prime+k_\mu^\prime k_\nu-g_{\mu\nu}k\cdot k^\prime+i\epsilon_{\mu\nu\alpha\beta}k^\alpha k^{\prime\beta},\nonumber\\
J^{\mu\nu}&=&\bar{\Sigma}\Sigma J^{\mu\dagger} J^\nu
\end{eqnarray}
and is calculated with the use of spin $\frac{3}{2}$ projection operator $P^{\mu\nu}$ defined as \[P^{\mu\nu}=\sum_{spins}\psi^\mu {\bar{\psi^\nu}}\] and is given by:
\begin{equation}
P^{\mu\nu}=-\frac{\not{p^\prime}+M^\prime}{2M^\prime}\left(g^{\mu\nu}-\frac{2}{3}\frac{p^{\prime\mu} p^{\prime\nu}}{M^{\prime 2}}+\frac{1}{3}\frac{p^{\prime\mu} \gamma^\nu-p^{\prime\nu} \gamma_\mu}{M^{\prime}}-\frac{1}{3}\gamma^\mu\gamma^\nu\right)
\end{equation}
In Eqn.(12), the decay width $\Gamma$ is taken to be an energy dependent P-wave decay width given by
\begin{eqnarray}
\Gamma(W)=\frac{1}{6\pi}\left(\frac{f_{\pi N\Delta}}{m_\pi}\right)^2\frac{M}{W}|{{\bf q}_{cm}|^3}\Theta(W-M-m_\pi)
\end{eqnarray}
where 
\[|{\bf q}_{cm}|=\frac{\sqrt{(W^2-m_\pi^2-M^2)^2-4m_\pi^2M^2}}{2W}\]
and $M$ is the mass of nucleon. The step function $\Theta$ denotes the fact that the width is zero for the invariant masses below the $N\pi$ threshold. ${|\bf q_{cm}|}$ is the pion momentum in the rest frame of the resonance.

\section{Weak Pion Production in Nuclei}
The reaction given in Eqns.4 and 5 taking place in the nucleus produces $\Delta^{++}$ or $\Delta^{+}$ which give rise to pions as decay product through $\Delta \rightarrow N \pi$ channel. In the incoherent pion production process only these pions are observed and no observation is made on other hadrons. However, in the nucleus there is a possibility, that the $\Delta$ produced in the reaction decays producing a pion such that the nucleus stays in the ground state. This leads to the process of coherent production of pions which are characterized by the forward production of pions in the direction of the lepton momentum transfer as the recoil of the nucleus is very small and negligible. 

In the following we discuss separately the incoherent and coherent production of pions in nuclei. 
\subsection{Incoherent Weak Production of Pions}
  When the reaction given by Eqns.4 or 5 takes place in the nucleus, the neutrino interacts with the nucleon moving inside the nucleus of density $\rho(r)$ with its corresponding momentum $\vec{p}$ constrained to be below its Fermi momentum $k_{F_{n,p}}(r)=\left[3\pi^2\rho_{n,p}(r)\right]^\frac{1}{3}$, $\rho_n(r)$ and $\rho_p(r)$ are the neutron and proton nuclear densities. The produced $\Delta$'s have no such constraints on their momentum. These $\Delta's$ decay through various decay channels in the medium. The production of $\Delta$ in the nucleus, leads to pion as it decays into nucleon and pion through $\Delta \rightarrow N\pi$ channel. However, these nucleons have to be above the Fermi momentum $p_F$ of the nucleon in the nucleus thus inhibiting the decay as compared to the free decay of the $\Delta$ described by $\Gamma$ in Eqn.12. Therefore, in the nuclear medium the decay width $\Gamma$ in Eqn.15 is to be modified in the nuclear medium.
 
The modification of $\Gamma$ due to Pauli blocking of nucleus has been studied in detail in electromagnetic and strong interactions\cite{oset}. The modified $\Delta$ decay width i.e. $\tilde\Gamma$ is written as\cite{oset}: 
\begin{equation}
\tilde\Gamma=\frac{1}{6\pi}\left (\frac{f_{\pi N \Delta}}{m_{\pi}}\right )^{2}\frac{M|{\bf q}_{cm}|^{3}}{W} F(k_{F},E_{\Delta},k_{\Delta})\Theta(W-M-m_\pi) 
\end{equation}
where $F(k_{F},E_{\Delta},k_{\Delta})$ is the Pauli correction factor given by 
\begin{equation}
F(k_{F},E_{\Delta},k_{\Delta})= \frac{k_{\Delta}|{{\bf q}_{cm}}|+E_{\Delta}{E^\prime_p}_{cm}-E_{F}{W}}{2k_{\Delta}|{\bf q^\prime}_{cm}|} 
\end{equation}
where $k_{F}$ is the Fermi momentum, $E_F=\sqrt{M^2+k_F^2}$, $k_{\Delta}$ is the $\Delta$ momentum and  $E_\Delta=\sqrt{W+k_\Delta^2}$.

Moreover, in the nuclear medium there are additional decay channels now open due to two body and three body absorption processes like $\Delta N \rightarrow N N$ and $\Delta N N\rightarrow N N N$ through which $\Delta's$ disappear in the nuclear medium without producing a pion while a two body $\Delta$ absorption process like $\Delta N  \rightarrow \pi N N$ gives rise to some more pions. These nuclear medium effects on $\Delta$ propagation are included by using a $\Delta$ propagator in which the $\Delta$ propagator is written in terms of $\Delta$ self energy $\Sigma_\Delta$. This is done by using a modified mass and width of $\Delta$ in nuclear medium i.e.  $M_\Delta \rightarrow M_\Delta + Re\Sigma_\Delta$ and $\tilde\Gamma \rightarrow \tilde\Gamma - Im\Sigma_\Delta$. There are many calculations of $\Delta$ self energy $\Sigma_\Delta$ in the nuclear medium \cite{oset}-\cite{oset3} and we use the results of \cite{oset}, where the density dependence of real and imaginary parts of $\Sigma_\Delta$ are parametrized in the following form:
\begin{eqnarray}
Re{\Sigma}_{\Delta}&=&40 \frac{\rho}{\rho_{0}}MeV ~~\mbox{and}\nonumber \\
-Im{{\Sigma}_{\Delta}}&=&C_{Q}\left (\frac{\rho}{{\rho}_{0}}\right )^{\alpha}+C_{A2}\left (\frac{\rho}{{\rho}_{0}}\right )^{\beta}+C_{A3}\left (\frac{\rho}{{\rho}_{0}}\right )^{\gamma}
\end{eqnarray}
In Eqn.18, the term with $C_{Q}$ accounts for the $\Delta N  \rightarrow \pi N N$ process, the term with $C_{A2}$ for two-body absorption process $\Delta N \rightarrow N N$ and the term with $C_{A3}$ for three-body absorption process $\Delta N N\rightarrow N N N$. The coefficients $C_{Q}$, $C_{A2}$, $C_{A3}$, $\alpha$, $\beta$ and $\gamma$ ($\gamma=2\beta$) are parametrized in the range $80<T_{\pi}<320MeV$ (where $T_{\pi}$ is the pion kinetic energy) as \cite{oset}
\begin{eqnarray}
C_i(T_{\pi})&=&ax^{2}+bx+c,~ \mbox{for}~ x=\frac{T_{\pi}}{m_{\pi}} 
\end{eqnarray}
The values of coefficients $a$, $b$ and $c$ are given in Table-I. taken from ref.\cite{oset}.
\begin{table}[h]
\caption{Coefficients of Eqns.18 and 19 for an analytical interpolation of $Im\Sigma_\Delta$}
\begin{tabular}{cccccc}
&$C_Q$(MeV) & $C_{A2}$(MeV)& $C_{A3}$(MeV) & $\alpha$&$\beta$ \\ \hline
 a & -5.19 & 1.06&-13.46&0.382&-0.038 \\
 b & 15.35&-6.64&46.17&-1.322&0.204\\
 c & 2.06&22.66&-20.34&1.466&0.613 \\
\end{tabular}
\end{table}

With these modifications, which incorporate the various nuclear medium effects on $\Delta$ propagation, the cross section for pion production on proton target is
\begin{eqnarray}
\sigma=\int \int \frac{d{\bf r}}{8\pi^3}\frac{d\bf{k^\prime}}{E_\nu E_l}\frac{1}{MM^\prime}
\frac{\frac{\tilde\Gamma}{2}-Im\Sigma_\Delta}{(W- M^\prime-Re\Sigma_\Delta)^2+(\frac{\tilde\Gamma}{2.}-Im\Sigma_\Delta)^2}\rho_p({\bf r})L_{\mu\nu}J^{\mu\nu}
\end{eqnarray}
For pion production on neutron target, $\rho_p({\bf r})$ in the above expression is replaced by $\frac{1}{3}\rho_n({\bf r})$, where the factor $\frac{1}{3}$ with $\rho_n$ comes due to suppression of $\pi^+$ production from neutron target as compared to the $\pi^+$ production from the proton target.

Therefore, the cross section for $\pi^+$ production in the neutrino interaction with the nucleus is given by
\begin{eqnarray}
\sigma=\int \int \frac{d{\bf r}}{8\pi^3}\frac{d\bf{k^\prime}}{E_\nu E_l}\frac{1}{MM^\prime}
\frac{\frac{\tilde\Gamma}{2}-Im\Sigma_\Delta}{(W- M^\prime-Re\Sigma_\Delta)^2+(\frac{\tilde\Gamma}{2.}-Im\Sigma_\Delta)^2}\left[\rho_p({\bf r})+\frac{1}{3}\rho_n({\bf r})\right]L_{\mu\nu}J^{\mu\nu}
\end{eqnarray}
 In case of antineutrino reactions the role of $\rho_p({\bf r})$ and $\rho_n({\bf r})$ are interchanged and $\rho_p + \frac{1}{3}\rho_n$ in the above expression is replaced by $\rho_n + \frac{1}{3}\rho_p$ and $L_{\mu\nu}^A$ is replaced by $-L_{\mu\nu}^A$ in Eqn 13.

\subsection{Results}
For our numerical calculation to evaluate lepton momentum and angular distribution as well as the total scattering cross section we use $\rho_n(r)=\frac{A-Z}{A}\rho(r)$ and $\rho_p(r)=\frac{Z}{A}\rho(r)$ in Eqn.(21), where the $\rho(r)$ is the nuclear density which we take to be a three parameter Fermi(3pF) for $^{12}{C}$ and $^{16}{O}$ nuclei given by
\begin{equation}
\rho(r)=\frac{\rho_0(1+w\frac{r^2}{\alpha^2})}{(1+exp((r-\alpha)/a))},
\end{equation}
where $\alpha$=2.355fm , $a=0.5224fm$ and $w=-0.149$ for $^{12}{C}$ nucleus and $\alpha$=2.608fm , $a=0.513fm$ and $w=-0.051$ for $^{16}{O}$ nucleus and a three parameter Gaussian(3pG) density for $^{56}{Fe}$ nucleus given by
\begin{equation}
\rho(r)=\frac{\rho_0(1+w\frac{r^2}{\alpha^2})}{(1+exp((r^2-\alpha^2)/a^2))},
\end{equation}
where $\alpha$=3.475fm , $a=2.33fm$ and $w=0.401$ for $^{56}{Fe}$ nucleus. The parameters are taken from de Jager et al.\cite{vries}.

In Fig.2, we present the results for the lepton momentum distribution $\frac{d\sigma}{dp_l}$ in $^{16}O$ as a function of lepton momentum at a fixed neutrino energy $E_\nu=1GeV$. The results for the momentum distribution without the nuclear medium effects have been shown by solid lines and with the nuclear medium effects have been shown by dashed lines. We find a reduction of about $15-30\%$ around the peak region of the momentum distribution. Further we find that around $80-85\%$ of these $\Delta's$ produce pions and the rest of them produce particle hole excitations. This is calculated by using $\frac{\tilde\Gamma}{2}$ and $C_3$ term in $Im\Sigma_\Delta$ for the production of pion and for medium absorption of $\Delta$, $C_1$ and $C_2$ terms in $Im\Sigma_\Delta$ in Eqn.(21). Therefore, the total effect of the medium modification is a reduction of $\approx 30-35\%$ in the peak region. However, pions once produced inside the nucleus, rescatter and some of them may be absorbed while coming out of the nucleus. We have estimated this absorption effect in an Eikonal arrroximation using the energy dependent mean free path of pions taken from Ref.\cite{oset}. We find a further reduction in the momentum distribution of the muon spectrum to be around $15-20\%$ in the peak region. In Fig.3, the results for the angular distribution $\frac{d\sigma}{dcos_\theta}$, where $\theta$ is the angle between the outgoing lepton and the neutrino, has been shown for neutrino energy $E_\nu=1GeV$. We find a reduction of around $10 \%$ in the forward direction when nuclear medium modification effects are encorporated. When the pion absorption effects are taken into account there is a further reduction of about $10\%$ in the forward direction. 

In Fig.4, we have shown the effects of medium modification and pion absorption for the total scattering cross section $\sigma$ in $^{16}O$. The solid line is the result for the cross section without nuclear medium effects, the dashed line shows the effect of medium modification and dotted line shows the effect when pion absorption is also taken into account. We find that the effect of medium modification results in a reduction of the cross section of around $10\%$ at $E_\nu=1GeV$ which further decreases with the increase in the neutrino energy. When pion absorption effects are taken into account there is a further reduction of around $10\%$ at $E_\nu=1GeV$. In Fig.5, we have shown the results with the nuclear medium modification and final state pion interaction effects for the total scattering cross section $\sigma$ in $^{12}C$(dotted line), $^{16}O$(dashed line) and $^{56}Fe$(solid line).

\section{Coherent Weak Pion Production}

The coherent pion production is the process in which the nucleus remains in the ground state. The study of such processes in electromagnetic interactions has shown that these are also dominated by $\Delta$-excitation in the intermediate energy region. Therefore, the weak pion production is also calculated using $\Delta$ dominance model. In this energy region the coherent pion production induced by neutrinos has been studied by Kelkar et al.\cite{kelkar} in a non-relativistic approach and by Kim et al.\cite{kim} in a relativistic mean field theory of the nucleus. At very high energies, the coherent pion production has been studied using PCAC and the Adler's theorem for forward production and extrapolating the results to non-zero $Q^2$ \cite{belkov}. Recently these calculations have been updated by Paschos and Kartavtsev\cite{paschos1} using a generalized PCAC.

We calculate the coherent pion production induced by charged current interaction i.e. $\nu (\bar\nu) + _Z^AX \rightarrow l^-(l^+) + \pi^+ +  _Z^AX$. The calculations are done in a local density approximation using $\Delta$ dominance in which the diagrams shown in Fig 6., contribute. 

The matrix elements corresponding to the Feynman diagrams shown in Fig.6 is given by
\begin{equation}
T=\frac{G_F}{\sqrt{2}}\bar{u}(k^\prime)\gamma^\mu(1-\gamma_5)u(k)~(J^\mu_{direct}+J^\mu_{exchange}) F{\bf (q - p_\pi)}
\end{equation}
where
\begin{equation}
{J^\mu_{direct}}=\sqrt{3}\frac{G_F}{\sqrt{2}}\cos\theta_C\frac{f_{{\pi} N {\Delta}}}{m_{\pi}}t_\sigma^\pi\sum_{s} {\bar {\Psi}}^s(p^\prime) \Delta^{\sigma \lambda}O_{\lambda \mu} \Psi^s(p)
\end{equation}
\begin{equation}
{J^\mu_{exchange}}=\sqrt{3}\frac{G_F}{\sqrt{2}}\cos\theta_C\frac{f_{{\pi} N {\Delta}}}{m_{\pi}}\sum_{s} {\bar {\Psi}}^s(p^\prime)t_{\sigma}^\pi O^{\sigma \lambda }\Delta_{\lambda \mu} \Psi^s(p) 
\end{equation}
\begin{equation}
F{\bf (q - p_\pi)}=\int d^3r(\rho_p({\bf r})+\frac{1}{3}\rho_n({\bf r}))e^{i\bf (q - p_\pi).r}
\end{equation}
$\Delta^{\sigma \lambda}$ or $\Delta_{\lambda \mu}$ is the relativistic $\Delta$ propagator given by Eqn.(14) and $O_{\lambda \mu}$ or $ O^{\sigma \lambda}$ is the weak $N-\Delta$ transition vertex given as the sum of $V^\mu + A^\mu$ using Eqns.7 and 8.

Using these expressions the following form of the double differential cross section for pion production is obtained

\begin{eqnarray}
\frac{d^2\sigma}{d\Omega_{\pi}dE_{\pi}}=\frac{1}{8}\frac{1}{(2\pi)^5}\frac{(E_\nu-E_\pi)}{E_\nu}|{\bf p_\pi}|\sum|T|^2
\end{eqnarray} 

where $|T|^2$ is obtained by squaring the terms given in Eqn.24. Similar expressions are derived for the lepton differential spectrum for the process induced by the charged currents.

The numerical evaluation of the cross section given in Eq.28 has been done for $^{12}{C}$, $^{16}{O}$ and $^{56}{Fe}$ using nuclear densities from Eqns.22 and 23 and the results are presented in Fig.7 to Fig.10. It is found that for coherent production the axial vector current gives dominant contribution and the contribution from the vector current is almost negligible. Therefore, the results for the neutrino and antineutrino reactions are almost similar. This is in agreement with the results obtained by Kim et al.\cite{kim} and Kelkar et al.\cite{kelkar}. 

In Fig.7, we present the results for the momentum distribution $\frac{d\sigma}{dp_\pi}$ in $^{16}O$ as a function of pion momentum at a fixed neutrino energy $E_\nu=1GeV$. The results for the momentum distribution without the nuclear medium effects have been shown by solid lines and with the nuclear medium effects have been shown by dashed lines. We find a reduction of about $60-80\%$ around the peak region of the momentum distribution. We find a further reduction in the pion momentum distribution to be around $15-30\%$ in the peak region due to pion absorption effects as shown by dotted lines in this figure. In Fig.8, the results for the angular distribution $\frac{d\sigma}{dcos_\pi}$ has been shown for neutrino energy $E_\nu=1GeV$. The angular distribution has been found to be very sharply peaked in the forward direction. The results without(with) medium modification effects have been shown by solid line(dashed line). We find a reduction of around $15-20\%$ in the forward direction when nuclear medium modification effects are encorporated. When the pion absorption effects are taken into account, which has been shown here by dotted line, there is a further reduction of about $8-10\%$ in the forward direction.

In Fig.9, we have presented the results of the total scattering cross section $\sigma(E_\nu)$ for the charged current coherent process in $^{16}O$. The result of $\sigma(E_\nu)$ without the nuclear medium effects has been shown by solid line and with the nuclear medium effects has been shown by dashed line. We find the reduction in the cross section because of the medium modification on $\Delta$ to be quite substantial at low energies. For example at $E_\nu=0.7GeV$, it is $\approx 40\%$ which decareses with the increase in energy and this reduction becomes $30\%$ at $E_\nu=1GeV$ and $20\%$ at $E_\nu=1.5GeV$. When pion absorption effects are taken into account, there is a further reduction of $25-30\%$ at these energies. In Fig.10 we show the results for the total cross section $\sigma(E_\nu)$ for coherent pion production in nuclei like $^{12}{C}$(dotted line), $^{16}{O}$(dashed line) and $^{56}{Fe}$(solid line).

\section{Conclusions}
We have calculated the weak pion production induced by charged currents in neutrino resctions at intermediate neutrino energies of $E_\nu \approx 1GeV$. This is relevant for the neutrino oscillation experiments being done with atmospheric neutrinos and the accelerator neutrinos at K2K and MiniBooNE.

The incoherent and coherent pion production have been calculated for $^{12}C$, $^{16}O$ and $^{56}Fe$ and the numerical results have been presented. The nuclear medium effects have been calculated using the renormalisation of $\Delta$ properties in the nuclear medium and the pion distortion effects using the energy dependent mean free path of pions as determined from the electromagnetic and strong interaction processes in this energy region.  The numerical results for the momentum distribution, angular distribution for lepton and/or pions and the total cross section have been presented and discussed. 
\acknowledgements
This work was financially supported by the Department of Science and Technology, Govt. of India under grant number DST SP/S2/K-07/2000. One of the authors(SA) would like to thank CSIR for the financial support. 
\begin{figure}
\includegraphics{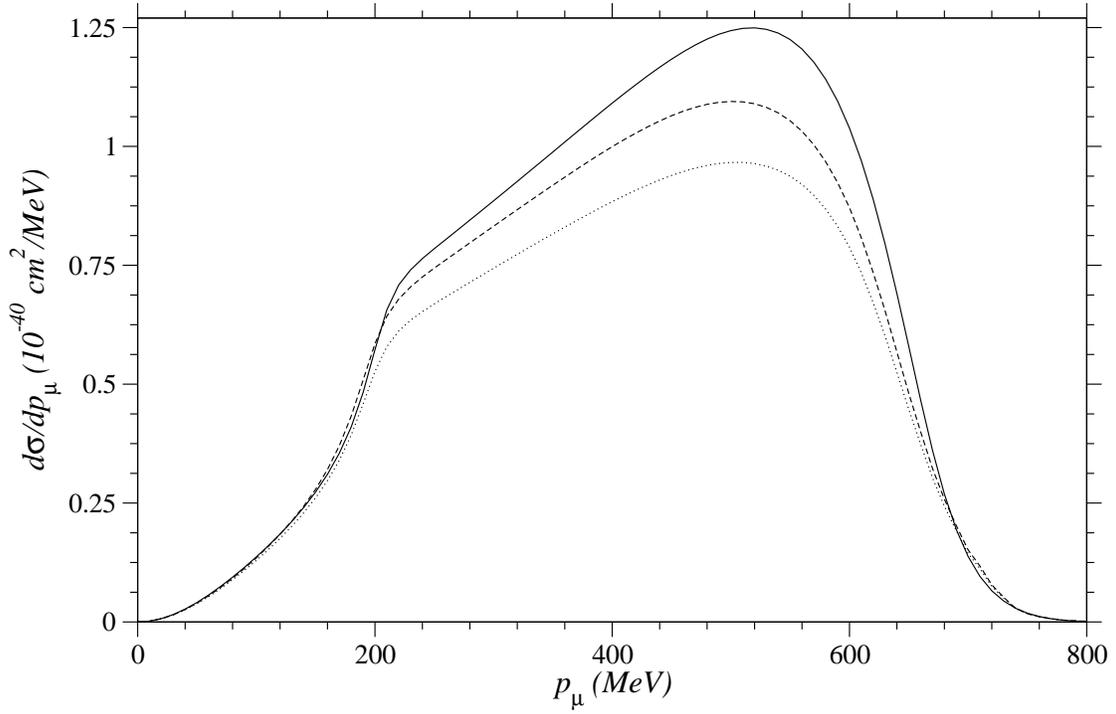}
\caption{$\frac{d\sigma}{dp_l}$ vs $p_l$ at $E_\nu=1.0 GeV$ for the charged current lepton production in oxygen with (dashed line) and without (solid line) nuclear medium effects. The result with nuclear medium and pion absorption effects is shown by dotted line.} 
\end{figure}

\begin{figure}
\includegraphics{pr2b.eps}
\caption{$\frac{d\sigma}{dcos_l}$ vs $ cos{\theta_l}$ at $E_\nu=1.0 GeV$ for the charged current lepton production in oxygen with (dashed line) and without (solid line) nuclear medium effects. The result with nuclear medium and pion absorption effects is shown by dotted line.}
\end{figure}

\begin{figure}
\includegraphics{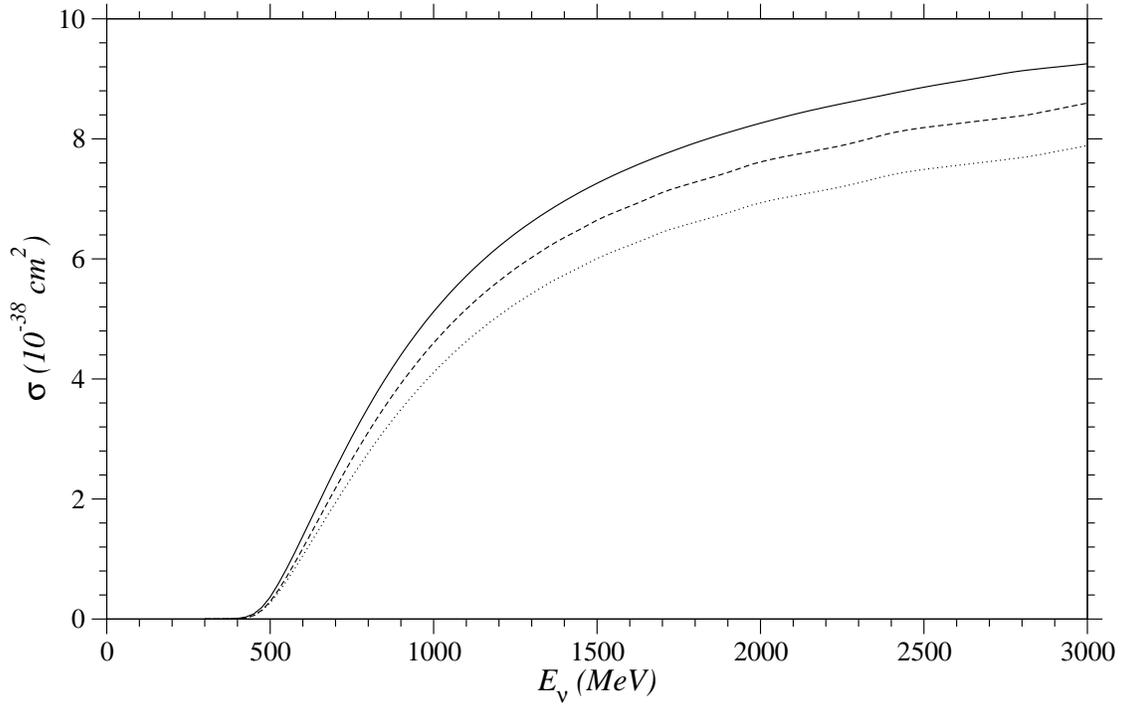}
\caption{Total scattering cross section $\sigma(E_\nu)$ for the charged current lepton production in $^{16}{O}$ nucleus without(solid line)  and with(dashed line) the nuclear medium effects. The result with nuclear medium and pion absorption effects is shown by dotted line.} \end{figure}

\begin{figure}
\includegraphics{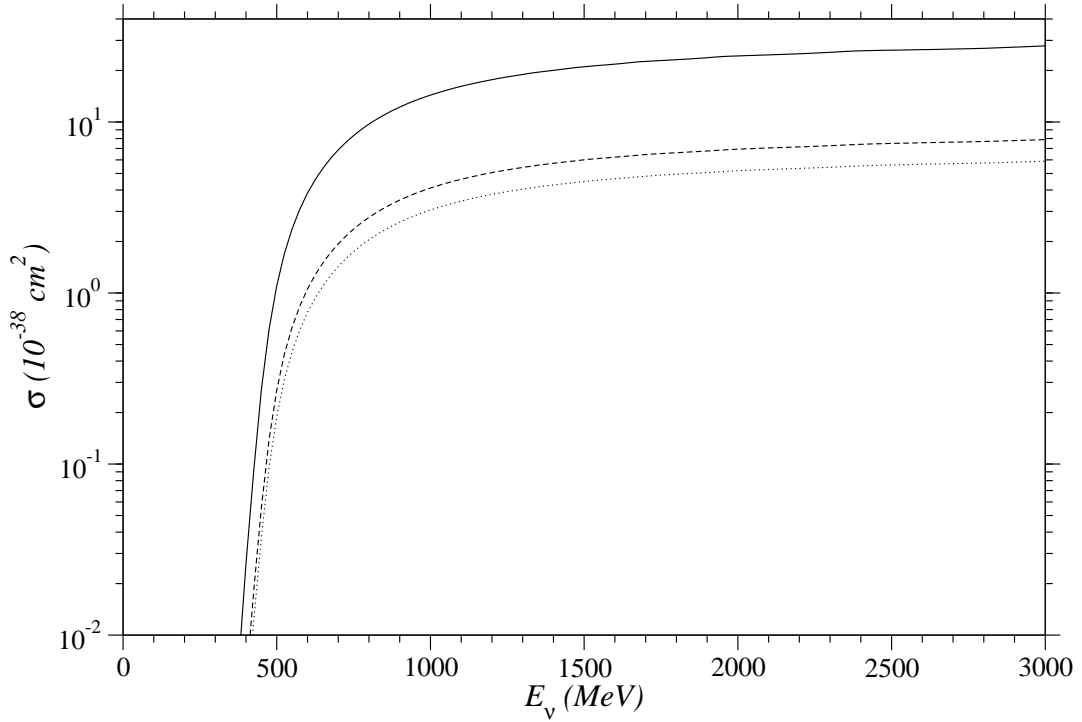}
\caption{Total scattering cross section $\sigma(E_\nu)$ for the charged current lepton production in $^{12}{C}$(dotted line), $^{16}{O}$(dashed line) and $^{56}{Fe}$(solid line) with nuclear medium and pion absorption effects.}
\end{figure}

\begin{figure}
\includegraphics{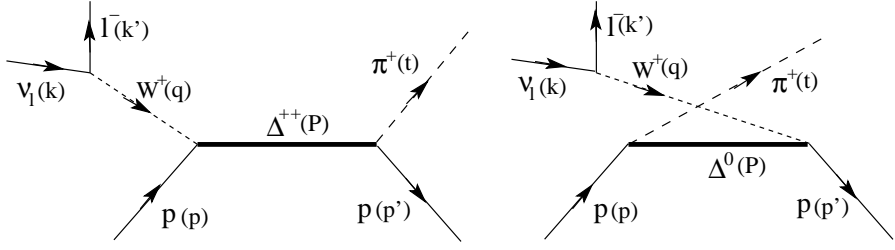}
\caption{Direct and exchange diagrams for the reaction $\nu_l(k)+ p(p)\rightarrow l^{-}(k^\prime)+ p(p^\prime) + \pi^+(t)$.}
\end{figure}

\begin{figure}
\includegraphics{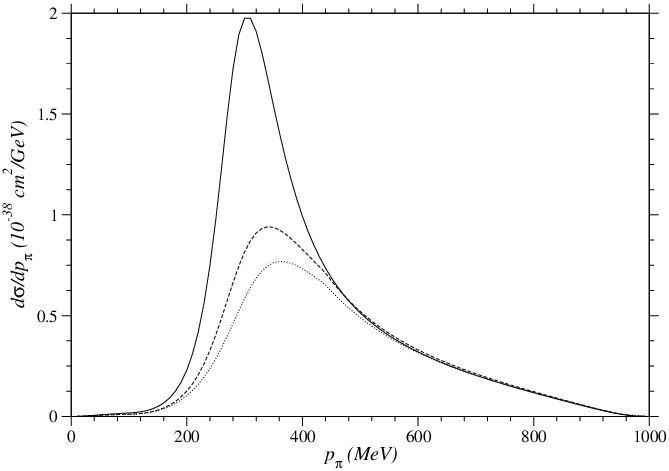}
\caption{$\frac{d\sigma}{dp_\pi}$ vs $p_\pi$ at $E_\nu=1.0 GeV$ for the charged current coherent pion production in oxygen with (dashed line) and without (solid line) nuclear medium effects. The result with nuclear medium and pion absorption effects is shown by dotted line.}
\end{figure}

\begin{figure}
\includegraphics{pr5b.eps}
\caption{$\frac{d\sigma}{dcos_\pi}$ vs $cos\theta_\pi$ at $E_\nu=1.0 GeV$ for the charged current coherent pion production in oxygen with (dashed line) and without (solid line) nuclear medium effects. The result with nuclear medium and pion absorption effects is shown by dotted line.}
\end{figure}

\begin{figure}
\includegraphics{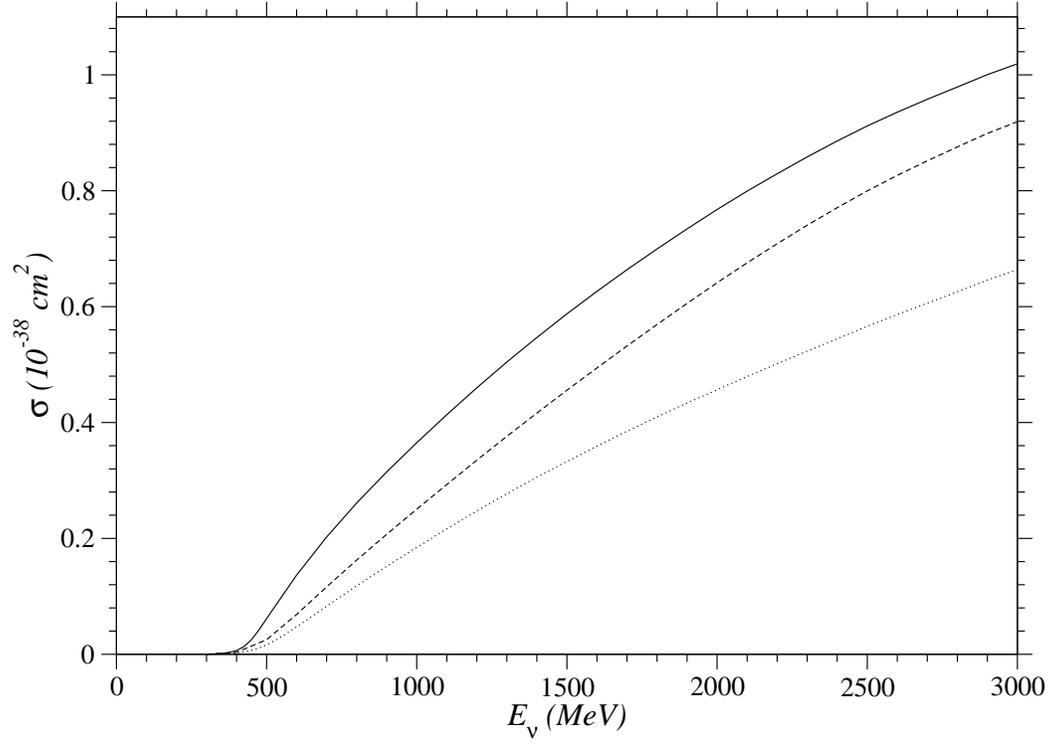}
\caption{Total cross section $\sigma(E_\nu)$ for the charged current coherent pion production in $^{16}{O}$ nucleus without(solid line) and with(dashed line) nuclear medium effects. The result with nuclear medium and pion absorption effects is shown by dotted line.}
\end{figure}

\begin{figure}
\includegraphics{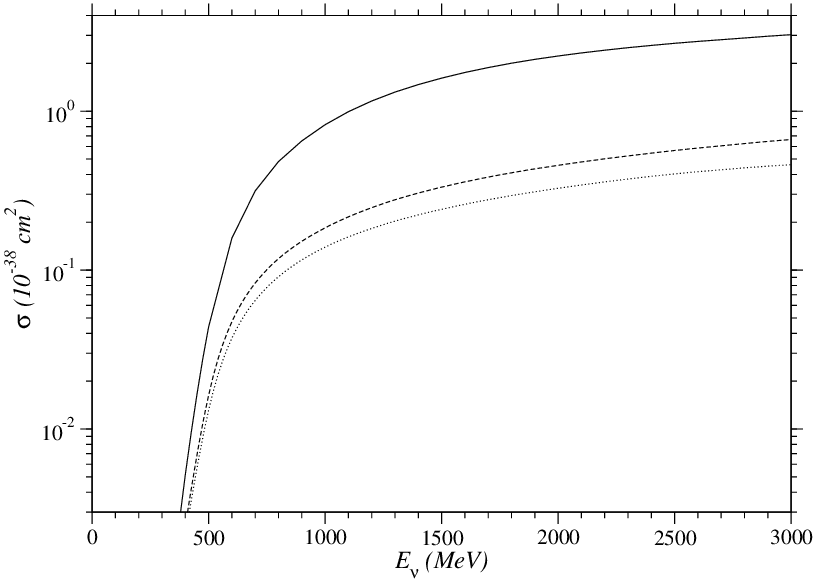}
\caption{Total scattering cross section $\sigma(E_\nu)$ for the charged current coherent pion production in $^{12}{C}$(dotted line), $^{16}{O}$(dashed line) and $^{56}{Fe}$(solid line) with nuclear medium and pion absorption effects.}
\end{figure}

\end{document}